\documentclass[AMA,Times1COL]{template_ARXIV2} %STIX1COL,STIX2COL,STIXSMALL

\articletype{Research Article}%

\received{Date Month Year}
\revised{Date Month Year}
\accepted{Date Month Year}
\journal{Journal}
\volume{00}
\copyyear{2024}
\startpage{1}

\usepackage{svg}
\usepackage{float}  %% added for figures to show up correctly
\usepackage{bbm}
\usepackage{changes}
\usepackage{graphicx}

\newcommand*{\LargerCdot}{\raisebox{-0.25ex}{\scalebox{1.2}{$\cdot$}}}
\DeclareMathOperator{\E}{E}
\DeclareMathOperator{\RR}{RR}

\begin{document}

\title{Simulation-based Bayesian predictive probability of success for interim monitoring of clinical trials with competing event data: two case studies}

\author[1]{Chiara Micoli}
\author[1]{Alessio Crippa}
\author[2,3]{Jason T. Connor}
\author[]{I-SPY COVID Consortium}
\author[1,4]{Martin Eklund}
\author[1,4]{Andrea Discacciati}

\authormark{MICOLI \textsc{et al.}}
\titlemark{Predictive probability of success with competing event data}

\address[1]{\orgdiv{Department of Medical Epidemiology and Biostatistics}, \orgname{Karolinska Institutet}, \orgaddress{\state{Stockholm}, \country{Sweden}}}

\address[2]{\orgdiv{ConﬂuenceStat LLC}, \orgname{Cooper City}, \orgaddress{\state{Florida}, \country{USA}}}

\address[3]{\orgdiv{University of Central Florida College of Medicine}, \orgname{Orlando}, \orgaddress{\state{Florida}, \country{USA}}}

\address[4]{These authors contributed equally to the work.}

\corres{Chiara Micoli, Department of Medical Epidemiology and Biostatistics, Karolinska Institutet, PO Box 281, SE-17177 Stockholm, Sweden. \email{chiara.micoli@ki.se}}	

% \presentaddress{This is sample for present address text this is sample for present address text.}

%\fundingInfo{Text}
%\JELinfo{ejlje}

\abstract[Abstract]{Bayesian predictive probabilities of success (PPoS) use interim trial data to calculate the probability of trial success.  These quantities can be used to optimize trial size or to stop for futility.
%Bayesian predictive probabilities of success (PPoS) analysis for interim monitoring of clinical trials enables the prediction of the probability of trial success at the conclusion of the trial given interim data.
In this paper, we describe a simulation-based approach to compute the PPoS for clinical trials with competing event data, for which no specific methodology is currently available. The proposed procedure hinges on modelling the joint distribution of time to event and event type by specifying Bayesian models for the cause-specific hazards of all event types. This allows the prediction of outcome data at the conclusion of the trial. The PPoS is obtained by numerically averaging the probability of success evaluated at fixed parameter values over the posterior distribution of the parameters. Our work is motivated by two randomised clinical trials: the I-SPY COVID phase II trial for the treatment of severe COVID-19 (NCT04488081) and the STHLM3 prostate cancer diagnostic trial (ISRCTN84445406), both of which are characterised by competing event data. We present different modelling alternatives for the joint distribution of time to event and event type and show how the choice of the prior distributions can be used to assess the PPoS under different scenarios. The role of the PPoS analyses in the decision making process for these two trials is also discussed.}


%\jnlcitation{\cname{%
%\author{Taylor M.},
%\author{Lauritzen P},
%\author{Erath C}, and
%\author{Mittal R}}.
%\ctitle{On simplifying ‘incremental remap’-based transport schemes.} \cjournal{\it J Comput Phys.} \cvol{2021;00(00):1--18}.}

\maketitle

\renewcommand\thefootnote{}
\footnotetext{\textbf{Abbreviations:} DMC,  data monitoring committee; HR, hazard ratio; PCa, prostate cancer; PCM, prostate cancer  mortality; PCH, piecewise constant cause-specific hazard; PPoS, predictive probability of success;  PSA, prostate-specific antigen; RR, risk ratio.}

\renewcommand\thefootnote{\fnsymbol{footnote}}
\setcounter{footnote}{1}

\section{Introduction}\label{sec1}
The interim monitoring of clinical trials has become an essential component of trial design\cite{jennisonStatisticalApproachesInterim1990}. Bayesian methods are particularly well suited for this purpose, as they enable the prediction of the probability of trial success at its conclusion based on interim data, averaged over the posterior distribution of the treatment effect and other nuisance parameters\cite{spiegelhalterMonitoringClinicalTrials1986}. By predicting the probability that the trial will provide sufficient evidence to support the study's primary hypothesis if the trial continues to its end, the Bayesian predictive probability of success (PPoS) can inform decisions regarding the continuation, modification, or early termination of the trial. If the PPoS is too low, and the trial is therefore predicted to be unlikely to achieve its objective, the trial may be stopped early for futility \cite{savilleUtilityBayesianPredictive2014, connorBayesianAdaptiveTrials2013}, reducing the risk of exposing patients to ineffective or harmful therapies and conserving resources that can be reallocated to more promising trials \cite{jennisonStatisticalApproachesInterim1990}. Conversely, if completing the prespecified follow-up period for patients who have already been enrolled is predicted to lead to success with a sufficiently high probability, enrollment of new patients can be stopped, and the primary outcome can be analysed at the conclusion of the follow-up period \cite{savilleUtilityBayesianPredictive2014, whiteCardiovascularSafetyStudy2012}. Furthermore, the PPoS can be employed to adjust the sample size at interim analyses\cite{broglioNotTooBig2014b} or to inform the timing of the primary analysis, by determining the minimum duration of the follow-up period required to achieve a certain PPoS \cite{choiEarlyDecisionClinical1985}. 

The PPoS has been primarily employed to monitor clinical trials with continuous or binary endpoints, e.g. in the I-SPY 2 trial \cite{parkAdaptiveRandomizationNeratinib2016} and the REST trial \cite{mcnameeEffectLowerTidal2021}. The use of PPoS in clinical trials with survival data is less common\cite{wilberComparisonAntiarrhythmicDrug2010, wangEvaluatingUtilizingProbability2013} and the  existing methodology is limited to the setting where the endpoint is the time span from some time origin (e.g. randomisation) until the occurrence of one single type of event \cite{broglioNotTooBig2014b, wangEvaluatingUtilizingProbability2013, yinBayesianAdaptiveRandomization2018, berryBayesianAdaptiveMethods2010, marion_predictive_2024, rufibachSequentiallyUpdatingLikelihood2016}. Competing events occur frequently in survival data from clinical trials, where patients may experience different types of events and the occurrence of one event precludes the occurrence of all other event types \cite{austinAccountingCompetingRisks2024}. For example, in randomised clinical trials evaluating potential treatments for hospitalised patients diagnosed with COVID-19, hospital discharge due to recovery and death without prior recovery are mutually exclusive (i.e. competing) events \cite{doddEndpointsRandomizedControlled2020}. 

To fill this gap, in this paper we take the approach of computing PPoS using statistical simulation \cite{trzaskomaPredictiveProbabilitySuccess2007} and extend it to the setting of survival data with competing events. We propose to model the joint distribution of time to event and event type by specifying Bayesian models for the cause-specific hazard functions, allowing the prediction of the outcome data at the conclusion of the trial and thus the computation of the PPoS. Our suggested procedure provides a general approach that can be used for interim monitoring of newly designed trials or to extend existing designs, for example for phase II \cite{yinPhaseIITrial2012} or phase III clinical trials\cite{broglioNotTooBig2014b}, to competing event data. 

Our work is motivated by two randomised clinical trials: the I-SPY COVID phase II trial for the treatment of severe COVID-19 \cite{filesISPYCOVIDAdaptive2022} (NCT04488081) and the STHLM3 prostate cancer diagnostic trial \cite{gronbergProstateCancerScreening2015f} (ISRCTN84445406). Both trials are characterised by a time-to-event outcome, namely time to recovery from COVID and time to prostate cancer mortality. In the I-SPY COVID trial, death from any cause is a competing event and the PPoS was used at interim monitoring to inform the decision to continue randomisation of patients to a specific investigational agent. In the STHLM3 trial, death from causes other than prostate cancer is a competing event and the PPoS was employed to predict the shortest follow-up period necessary to accumulate enough prostate cancer deaths in order to reach sufficient power for the final analysis.

\section{Methods}\label{sec}

\subsection{Notation and preliminaries}\label{notation}
For each subject $s$, where $s=1,\ldots,S$, let $T_s$ be the an absolutely continuous variable representing the true event time with support on the positive real half-line, and let $X_{s} \in \{1,\ldots,I\}$ be the competing event indicator. The right-censored event time is denoted by $Y_s = \min(T_s, C_s)$, where $C_s$ is the censoring time (generally assumed independent of $T_s$ and $X_s$), and $\Delta_s = \mathbbm{1}(T_s \leq C_s) \cdot X_s \in \{0,1,\ldots,I\}$ is the observed event indicator, where 0 indicates right censoring. The pair ($Y_s$, $\Delta_s$) defines the right-censored outcome data. Lastly, let $Z_{s}$ be a vector of baseline covariates, including at least the randomisation arm $A_s$. In total, we define the random vector $D_s = (Y_s, \Delta_s, Z_{s})$. Hereafter, we suppress the subscript $s$ to simplify notation. The cause-specific hazard for the $i$-th competing event is defined as

\begin{equation*}
\lambda_i(t) = \lim_{dt \downarrow 0} \frac{\Pr(T \leq t+dt, X = i \,|\, T > t)}{dt}.
\end{equation*}
The cause-specific hazards for all events collectively define the joint distribution of $(T, X)$ \cite{andersenCompetingRisksMultistate2002}. The crude cumulative probabilities (risks) of experiencing the competing event $i$ by the time $t$ are defined as $F_i(t) = \Pr(T \leq t, X = i) = \int_0^t \lambda_i(u) S(u)du$, where $S(t) = \Pr(T > t) = \exp(-\sum_{i=1}^{I} \int_0^t \lambda_i(u) du)$ is the event-free survival function. The cause-specific hazards and the crude cumulative risks are not only two possible quantities that define the joint distribution of $(T, X)$, but they are also quantities of scientific interest in clinical research \cite{austinAccountingCompetingRisks2017}. %In the remainder of this paper, quantities sampled from a distribution are denoted by a tilde.

At interim monitoring, the goal is to predict the probability of trial success at the final analysis given the interim data and the assumed prior distributions on the parameter vector $\theta$ for the data-generating process used to predict future data. Let $d_0$ be the observed data at the time of the interim analysis, $D_f$ be the future data available only at the time of the final analysis, which encompasses both observed and yet-to-be-observed data, and let $R$ be the critical region such that if $D_f \in R$ the trial will be declared a success. The PPoS, introduced by Spiegelhalter et al. \cite{spiegelhalterMonitoringClinicalTrials1986}, is defined as: 

\begin{equation*}
\textrm{PPoS} = \Pr(D_{f} \in R \, | \, d_0) =  \int_{\Theta} \Pr(D_f \in R\,|\,\theta) \; p(\theta\,|\,d_0)\, d\theta = \E_{\theta|d_0}[\E[\mathbbm{1}(D_f \in R)\,|\,\theta]].
\end{equation*}
In the PPoS, the conditional power $\Pr(D_f \in R\,|\,\theta)$, defined as the probability of trial success given the interim data and a fixed $\theta$, is averaged over the posterior distribution of $\theta$, denoted $p(\theta\,|\,d_0)$. The definition of trial success can be based on Bayesian posterior probabilities, as in the I-SPY COVID trial, or on frequentist p-values, as in the STHLM3 trial \cite{savilleUtilityBayesianPredictive2014}.

Three possible types of data must be considered when computing the PPoS at interim monitoring with time-to-event data. Firstly, there are completely observed data, where the event type, the time to the event, and the baseline covariates are known ($d_{obs} = (t, x, z)$). Secondly, there are partially observed data, where the baseline covariates are known, but the follow-up time is right-censored at the time of interim monitoring ($d_{cens} = (c, 0, z)$). Thirdly, there are completely unobserved data, which will be observed in the future if new patients are recruited and followed up ($D_{new} = (Y, \Delta, Z)$). In order to relate these types of data to the PPoS formula, note that $d_0$ is given by stacking $d_{obs}$ and $d_{cens}$ ($d_0 = (d_{obs}, d_{cens})$). The data $D_f$ include $d_{obs}$, $D_{new}$, but also $D_{cens} = (Y, \Delta, z)$, representing the updated data at the conclusion of the trial for the patients in $d_{cens}$ ($D_f = (d_{obs}, D_{cens}, D_{new})$). At interim monitoring, $D_{new}$ and $D_{cens}$ are not fully observed and to calculate the simulation-based PPoS they need to be predicted from statistical models. We return to the different types of data and how they are used in the calculation of the PPoS with competing risk data in Section \ref{PPoS}.

\subsection{Predictive probability of success with competing event data}\label{PPoS}

The framework to predict the probability of trial success via statistical simulation can be divided in three phases: modelling, prediction, and analysis \cite{trzaskomaPredictiveProbabilitySuccess2007}. The proposed approach specialises the first and second phase to allow the modelling and prediction of competing event data, which is then analysed in the third phase.

\bmsubsubsection*{Modeling phase:\\}
\begin{enumerate}
\item  Model the distribution for the time-to-event and event type conditional on baseline covariates $(T, X\,|\,Z=z)$ using Bayesian models and interim data $d_0$:
\begin{equation*}
p(t, x \, |\,z, \theta)
\end{equation*}  
We propose to do so by modelling the cause-specific hazards $\lambda_{ia} (t \, | \, z, \theta_{ia})$ for each event type $i = \{1, 2\}$ using Bayesian parametric models, stratifying over randomisation arm $a = \{0, 1\}$ as is common practice in other applied work\cite{broglioNotTooBig2014b, berryBayesianAdaptiveMethods2010}, and possibly conditioning on other baseline covariates.  Different parametric models can be assumed for the different combinations of event type and randomisation arm (see Section \ref{sec4} for an example). \\

\item If additional patients are to be enrolled in the trial, model the joint distribution $p(z|\eta)$ for the baseline covariates $Z$ using Bayesian models. The randomisation arm is generally independent of the other baseline covariates, which may simplify modelling. As the randomisation probabilities are known, a Bayesian model for randomisation arm is not necessary.
\end{enumerate}

The prior distributions $p(\theta_{ia})$ and $p(\eta)$ for the parameters of the models may be weakly informative, if the goal in the next phase is to obtain predictions that can be considered to be based on interim data only\cite{spiegelhalterMonitoringClinicalTrials1986}. Sceptical, optimistic, or more generally informative priors may be specified for example when conducting sensitivity analyses (see Section \ref{sec3} for an example). 

\bmsubsubsection*{Prediction phase:\\}
\begin{enumerate}
    \setcounter{enumi}{2}    
        \item Make a single draw, denoted $\tilde{\theta}_{ia}$ and $\tilde{\eta}$, from the posterior distributions $p(\theta_{ia} \, | \, d_0)$ and $p(\eta \, | \, d_0)$ of the models in step (1) and (2), respectively. \\
    
    \item  Simulate the baseline covariate data for the additional patients to be enrolled in the trial (if any). New values $\widetilde{Z}$ are sampled from the distribution $p(z \,| \, \tilde\eta)$. With known randomisation probabilities and a fixed total number of future patients, the number of patients in either randomisation arm can be sampled from a Binomial distribution. \\
        
    \item Predict time-to-event and event type $(\widetilde{T}, \widetilde{X}\,|\,Z=z)$ by sampling from the distribution:
\begin{equation*}    
p_a(t, x \,| \, z, \tilde\theta_{1a}, \tilde\theta_{2a}).
\end{equation*} In practice, this can be done in two steps, noting that the joint distributions can be decomposed as the product of two univariate distributions: $p_a(t \,| \, z, \tilde\theta_{1a}, \tilde\theta_{2a}) \cdot p_a(x \,| \, t, \: z, \tilde\theta_{1a}, \tilde\theta_{2a})$ \cite{beyersmannSimulatingCompetingRisks2009}. \\
    
    \begin{enumerate}
    \item The event times $\widetilde{T}$ are simulated from the distributions $p_a(t \,| \, z, \tilde\theta_{1a}, \tilde\theta_{2a})$, which are defined by the all-cause hazard functions $\lambda_{\LargerCdot{}a} (t \, | \, z, \tilde\theta_ {1a}, \tilde\theta_ {2a}) = \lambda_{1a} (t \, | \, z, \tilde\theta_ {1a}) + \lambda_{2a} (t \, | \, z,  \tilde\theta_{2a})$. For those patients already in the trial and censored at the interim analysis ($d_{cens}$), their updated event times are simulated conditional on being still at risk at their censoring time $c$. This is done by treating the censoring time as a left-truncation time point and consequently by simulating times from the distribution $\frac{p_a(t \,| \, z, \tilde\theta_{1a}, \tilde\theta_{2a})\cdot \mathbbm{1}(t>c)}{1-F_a(c \,| \, z, \tilde\theta_{1a}, \tilde\theta_{2a})}$, where $F$ is the cumulative distribution function of $T$. \\ 
   
    \item The event types $\widetilde{X}$ are simulated from Bernoulli experiments, where the conditional probability \\
    $\Pr(X = 1 \, | \, t, z, \tilde\theta_{1a}, \tilde\theta_{2a}) = \frac{\lambda_{1a} \, (t \, | \, z, \tilde\theta_ {1a} )}{ \lambda_{\LargerCdot{}a} \, (t \, | \, z, \tilde\theta_ {1a}, \tilde\theta_ {2a} )}$. \\
\end{enumerate}    
If necessary, generate censoring times according to the study design (e.g., censoring at the maximum allowed follow-up) and derive the quantities $\widetilde{Y}$ and $\widetilde{\Delta}$ as described in Section \ref{notation}. \\

    \item Stack the observed and predicted data in $\widetilde{D}_f=(d_{obs}, \widetilde{D}_{new}, \widetilde{D}_{cens})$, where $\widetilde{D}_{new} = (\widetilde{Y}, \widetilde{\Delta}, \widetilde{Z})$ and  $\widetilde{D}_{cens} = (\widetilde{Y}, \widetilde{\Delta}, z)$
    \end{enumerate}

\bmsubsubsection*{Analysis phase:\\}
\begin{enumerate}
    \setcounter{enumi}{6}
    \item Analyse the data $\widetilde{D}_f$ using the frequentist or Bayesian analysis method $g(\cdot)$, which maps $\widetilde{D}_f$ to the critical region $R$. Evaluate whether the statistic(s) of interest $g(\widetilde{D}_f)$ falls within $R$, leading to trial success, or not:
        \begin{equation*}
        G = \mathbbm{1}(g(\widetilde D_{f}) \in R \, |\, \tilde\eta, \tilde\theta_{10}, \tilde\theta_{11}, \tilde\theta_{20}, \tilde\theta_{21}).
        \end{equation*}
\end{enumerate}

Repeat steps (3) to (7) a sufficiently large number of times ($K$) and record $G^{(k)}$ for each iteration $k=1, \ldots, K$. The proportion of  successful trials approximates the PPoS 
\cite{johnsUsePredictiveProbabilities1999, trzaskomaPredictiveProbabilitySuccess2007}, by effectively averaging the conditional probability of trial success over the posterior distribution:
    \begin{equation*}
    \textrm{PPoS} \approx  \frac{1}{K} \sum_{k=1}^{K} G^{(k)}.
    \end{equation*}

The methodology can be extended to more than two competing events by modelling the cause-specific hazards for all events in step (1) and simulating the event type from a multinomial distribution in step (5).

\subsection{Software}

All the analyses were performed in R version 4.3.2 \cite{Rversion}. Hazard models were fitted to the data in Stan version 2.32.6 using full Bayesian statistical inference with Monte Carlo Markov Chain (MCMC) sampling \cite{carpenterStanProbabilisticProgramming2017}. The \texttt{brms} package version 2.21.0 was used as the interface to Stan \cite{burknerBrmsPackageBayesian2017}. For Bayesian models with Weibull hazard, we used a custom response distribution available at \url{https://github.com/anddis/brms-weibullPH}. The non-parametric Aalen-Johansen estimates of the cumulative incidence functions were obtained using the \texttt{survival} package version 3.7-0 \cite{survival-package}.  We provide the R code for reproducing the analyses on a synthetic dataset at \url{https://github.com/cmicoli/PPoS-CompetingRisks}.

\section{Motivational Example 1: The I-SPY COVID trial} \label{sec3} 
\subsection{Trial description and motivation} \label{secISPY}
The I-SPY COVID (NCT04488081) is a multicentre phase-II open-label adaptive platform trial aimed at identifying potential therapeutics for severe COVID-19\cite{filesISPYCOVIDAdaptive2022}. Patients with a positive SARS-CoV-2 test by PCR or rapid antigen testing on $\geq6$ L/min oxygen or intubated were screened for eligibility. Patients meeting the eligibility criteria were randomly assigned to either a common control arm, consisting of a backbone regimen of dexamethasone and remdesivir, or to backbone plus one of up to 4 investigational agents in the trial at any one time. Patients who provided informed consent were included in the modified intention-to-treat (mITT) population. The maximum number of patients who could be randomised to each investigational arm was 125. 

The primary endpoints of the trial were time to durable recovery and time to death due to any cause. Time to recovery was defined as the time elapsed between randomisation and the earliest second consecutive day with WHO COVID level 4 (i.e. <6 L/min oxygen) or below. In the analyses, follow-up was censored at 60 days if patients were still at risk. An agent would graduate if the vector consisting of the posterior probability of the cause-specific hazard ratio (csHR) for recovery being >1 and the posterior probability of the HR for all-cause mortality being <1 was in the 2-dimensional critical region $[0.975, 1) \cup [0.900, 1)$. Separate Bayesian Weibull models with weakly informative priors were used to model the cause-specific hazard for recovery and the hazard for all-cause mortality in the mITT population and to derive the posterior probabilities of graduation. These posterior probabilities were updated every two weeks and reviewed by an independent data monitoring committee (DMC). The primary role of the DMC was to ensure the safety of the patients in the trial, but also to recommend to the principal investigators that an  agent should graduate or be dropped for futility, based on unblinded review of trial results. A detailed description of the study design, operations, and statistical analysis plan of the trial can be found elsewhere\cite{filesISPYCOVIDAdaptive2022, filesReportFirstSeven2023, calfeeClinicalTrialDesign2022}. The analysis presented here replicates that performed in one of the interim reviews, with minor differences in statistical software and data, and it focuses solely on the time-to-recovery endpoint. 

At the interim analysis approximately 15 weeks after the first patient was randomised to the backbone plus cyclosporine arm (2021-07-19), there were 58 patients in the cyclosporine arm and 75 in the control arm. A total of 34 patients in the cyclosporine arm and 47 in the control arm had recovered, while the corresponding numbers of patients who died before recovery were 7 and 12, respectively. The posterior median $\textrm{csHR}_{\textrm{recovery}}$ for cyclosporine versus control was 0.97 (95\% quantile credible interval: 0.63, 1.47), indicating that the instantaneous recovery rate was similar in the two arms. The corresponding posterior probability of graduation was 0.44, outside the critical region $R = [0.975, 1)$. Given that the interim data on the recovery endpoint was not in favour of cyclosporine after almost 50\% of the maximum allowed number of patients had been randomised to cyclosporine, the DMC asked to predict the probability that cyclosporine would demonstrate sufficient benefit to meet the graduation criterion at the maximum sample size. The PPoS analysis was not prespecified in the trial protocol. %The analysis reported here replicates that performed in 2021 with minor differences in statistical software and data and it focuses only on the time-to-recovery endpoint. 

\subsection{Results}

In the modelling phase, we modelled the joint distribution of $(T, X, W, A)$ as $p(t,x \,| \, w,a)\, p(w)\, p(a)$ using interim data $d_0$. The binary variable $W$ represents baseline WHO COVID level of 6 or 7 ($W=1$) versus 5 ($W=0$) and the binary variable $A$ represents cyclosporine ($A=1$) versus control arm ($A=0$). Due to randomisation, we assumed $W$ to be independent of $A$. The joint distribution $(T, X \, | \,A=a, W=w)$ was modelled via the cause-specific hazard functions for the two competing events, recovery from COVID ($i=1$) and death from any cause ($i=2$). In particular, we specified Bayesian Weibull models stratified by randomisation arm so as not to impose any proportionality constraint between the cause-specific hazards for the cyclosporine and the control arm: $\lambda_{ia}(t|w, \alpha_{ia}, \gamma_{ia}, \nu_{ia})= u_{ia}\nu_{ia}t^{\nu_{ia}-1}$. We let the parameter $u_{ia}$ depend on baseline WHO COVID level, such that $\log(u_{ia}) = \alpha_{ia} + \gamma_{ia}w$, while the parameter $\nu_{ia}$ did not depend on any covariates. The following priors completed the model specification: $\alpha_{ia} \sim \textrm{Normal}(0, 20)$, $\gamma_{ia} \sim \textrm{Normal}(0, \sqrt{0.5})$, $\nu_{ia} \sim \textrm{Exponential}(1)$. The baseline WHO COVID level $W$ followed a $\textrm{Bernoulli}(\eta)$ distribution with a noninformative prior $\eta \sim \textrm{Beta}(1, 1)$. 

In the prediction phase, we fixed the number of new patients in the cyclosporine arm to $125-58=67$, and given that the randomisation probability to the cyclosporine arm was 0.45, we sampled the number of patients in the control arm from a $\textrm{NegativeBinomial}(67,0.45)$ distribution. For these new patients, we simulated the  baseline WHO COVID level, time to event, and event type ($\widetilde{D}_{new}$). For the 33 patients still at risk at the interim analysis ($d_{cens}$), the outcome data were simulated conditional on the observed baseline WHO COVID level, randomisation arm, and their censoring times ($\widetilde{D}_{cens}$). Predicted follow-up times were censored if they exceeded 60 days. The interim and predicted data were stacked in the dataset $\widetilde{D}_f$.

In the analysis phase, we analysed interim and predicted data together using Bayesian proportional hazard models and then derived the posterior probability $\Pr(\textrm{csHR}_{\textrm{recovery}}>1.00 \,|\, \widetilde{D}_f)$ over 4000 draws, as prespecified in the SAP.

We repeated the prediction and analysis phase $K=2500$ times and approximated the PPoS as 
\begin{equation*}
\textrm{PPoS} \approx \frac{1}{2500}\sum_{k=1}^{2500} \mathbbm{1}(\Pr(\textrm{csHR}_{\textrm{recovery}}>1.00 \,|\, \widetilde{D}_f^{(k)}) \ge 0.975 \,|\,\tilde\theta^{(k)}) = 0.061,
\end{equation*} 
indicating that Cyclosporine would not likely have shown a benefit on the COVID recovery rate even at maximum sample size. Combining the information from the prespecified posterior probability analysis and the PPoS, the DMC recommended stopping randomisation of patients to the cyclosporine arm.

As the PPoS was low, we assessed in a sensitivity analysis how more enthusiastic priors about the $\textrm{csHR}_{\textrm{recovery}}$ for cyclosporine versus control for the remainder of the trial would affect the PPoS\cite{spiegelhalterBayesianApproachesRandomized1994}. The rationale for this analysis was to reassure the DMC that their decision to stop the cyclosporine arm due to futility would be made even under optimistic assumptions about the effect of cyclosporine on the recovery rate. In the modelling phase of $(T, X \, | \, A=a, W=w)$, we forced the cause-specific hazards for the two randomisation arms to be proportional with proportionality constant $\exp(\beta_i)$ , when specifying $\log(u_{i}) = \alpha_{i} + \beta_{i}a + \gamma_{i}w$, $i=\{1, 2\}$, forcing the cause-specific hazards for the two randomisation arms to be proportional with proportionality constant $\exp(\beta_i)$. The priors for $\alpha_i$, $\gamma_i$, and $\nu_i$ were defined as in the main PPoS analysis, while $\beta_2 \sim \textrm{Normal}(0, \sqrt{0.5})$. We let the prior $\beta_1 \sim \textrm{Normal}(\mu, \sigma)$ take on different values for the mean and standard deviation, covering different scenarios ranging from weakly neutral to strongly optimistic priors. We explored 15 different scenarios by setting $\mu=\{\log(1), \log(1.2), \log(2), \log(3), \log(4)\}$ and $\sigma=\{\sqrt{0.1}, \sqrt{0.2}, \sqrt{0.5}\}$. Some of these priors were extremely optimistic, for example the equal-tailed  95\% prior limits for the $\textrm{csHR}_{\textrm{recovery}}$ under the $\textrm{Normal}(\log(4), \sqrt{0.1})$-prior were $(2.15, 7.43)$. These values were considerably larger than what was considered plausible at the time of the trial design, where the largest value used in the sample size calculations was 1.80.

\begin{figure}[h]
\centerline{\includegraphics[width=0.75\textwidth]{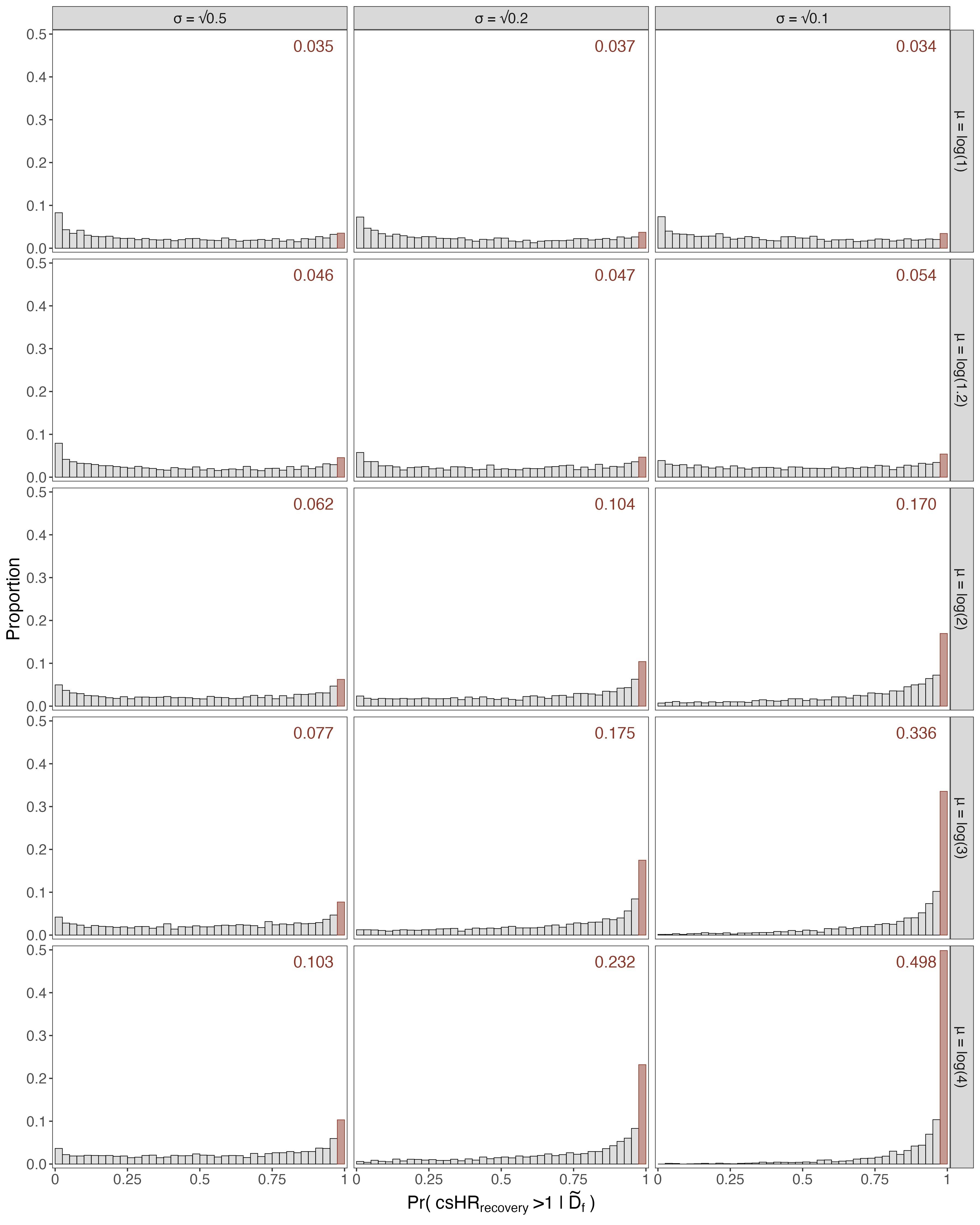}}
\caption{Distributions of the posterior probability of graduation $\Pr(\textrm{csHR}_{\textrm{recovery}}>1.00 \,|\, \widetilde{D}_f)$ over 2500 simulations for the 15 different scenarios considered in the sensitivity analysis for the I-SPY COVID trial. The red bar corresponds to the PPoS, i.e. to the proportion of simulations for which $\Pr(\textrm{csHR}_{\textrm{recovery}}>1.00 \,|\, \widetilde{D}_f) \geq 0.975$. Each panel corresponds to a different prior $\beta_1 \sim \textrm{Normal}(\mu, \sigma)$ specified in the modelling phase. The values specified for the prior's parameters $\mu$ and $\sigma$ are reported along the rows and columns, respectively. \label{fig1}}
\end{figure}

Figure \ref{fig1} shows the distribution of the posterior probability $\Pr(\textrm{csHR}_{\textrm{recovery}}>1.00 \,|\, \widetilde{D}_f)$ over 2500 simulations for the 15 different scenarios. The PPoS varied from 0.034 to 0.498, indicating that even an unrealistically optimistic and strong prior for $\textrm{csHR}_{\textrm{recovery}}$ resulted in a relatively low PPoS. This sensitivity analysis confirms that cyclosporine is unlikely to reach the graduation threshold for recovery at maximum sample size even under an extremely optimistic scenario.

\section{Motivational Example 2: Long-term follow-up of the STHLM3 trial} \label{sec4} 

\subsection{Trial description and motivation}
The population-based diagnostic STHLM3 trial (ISRCTN84445406) was conducted between May 2012 to January 2015. A random sample of men aged 50-69 years living in Stockholm, Sweden, were invited to participate in a one-time prostate cancer screening intervention using prostate-specific antigen (PSA) and the Stockholm3 test, a prediction model based on clinical variables, plasma protein biomarkers, and a polygenic risk score \cite{gronbergProstateCancerScreening2015f}. One of the secondary aims of the STHLM3 trial was to evaluate the efficacy of screening invitation on long-term prostate cancer mortality (PCM) in the presence of death from other causes as a competing event, which is the focus of the analysis reported here. The population eligible for random invitation to participate in the screening intervention consisted of  240,494 men. Of these, 169,621  (70\%) were randomly invited and 59,088  (35\%) participated. 

An interim analysis (IA) was performed using follow-up data until the end of 2018 to compare the crude cumulative risk of PCM between the two groups at 6.5 years after invitation\cite{micoliA0890Stockholm3Prostate2023}. Death from causes other than prostate cancer was considered a competing event. Cumulative risks were estimated using the Aalen-Johansen estimator and the difference in PCM risk was summarised by a risk ratio (RR) with the non-invited group as the referent. A p-value for the two-sided alternative hypothesis $H_1: \RR \neq 1$ versus the null $H_0: \RR = 1$ was calculated as $p=2\cdot\Phi\left(-\left| \frac{\log(\RR)}{ASE(\log(\RR))} \right|\right)$, where $ASE(\log(\RR))$ is the asymptotic standard error for the $\log(\RR)$ and $\Phi(\cdot)$ is the standard normal cumulative distribution function. We considered a two-sided alternative hypothesis because we could not rule out a priori that screening invitation might actually increase PCM risk. At interim analysis, the significance level $\alpha(6.5)$ was set to 0.02. During a maximum of 6.7 years of follow-up, we observed 101 prostate cancer deaths in the invited group and 54 in the non-invited group, while the estimated PCM risks were 8.6 and 13.2 per 10,000 men, respectively (Table \ref{tab:sthlm3-events}). The $\RR(6.5)$ for PCM was 0.66 (98\% confidence interval (CI): 0.41, 1.06), $p=0.04$, which favoured the invited group but did not reach formal statistical significance.  

\begin{table}[h]
\centering
\begin{tabular}{l|lllll}
 Randomisation group & N & Deaths from PCa & Deaths from other causes & PCa mortality risk & Other causes mortality risk\\ \hline
Invited     & 169,621 & 101  & 6644    & 8.6  per 10,000 men      & 525 per 10,000 men              \\
Not invited & 70,873   & 54  & 2814    & 13.2 per 10,000 men      & 536 per 10,000 men        \\ 
\hline
\end{tabular}
\caption{Interim analysis results of the STHLM3 trial. Number of randomised men (N), deaths from prostate cancer (PCa), deaths from other causes and corresponding mortality risks for the invited and non-invited randomisation groups. Risks the end of follow-up were estimated using the Aalen-Johansen estimator treating the other event as a competing event.} \label{tab:sthlm3-events} 
\end{table}

The time point for the final comparison of PCM risk between the two invitation arms was prespecified to be between 10 and 15 years after study initiation. The PPoS was used to guide the selection of the specific time point for the final analysis ($FA$). We first predicted the probability of finding a statistically significant RR after 10 years given the interim data and then we repeated the prediction process five other times, adding one year at a time until a maximum of 15 years, thus allowing for more prostate cancer deaths to accumulate. 
%\deleted{If the PPoS with 10 years of follow-up was considered sufficiently high, then it would be decided to conduct the final analysis at 10 years. Otherwise, a later time point (e.g. 12 or 15 years) with a sufficiently high PPoS would be chosen for the final analysis, allowing for more prostate cancer deaths to accumulate during the longer follow-up.} 
The alpha level for the final analysis $\alpha(FA)$ used in the PPoS estimation was set at 0.035 (critical region $R = (0, 0.035]$) so that that the overall type I error rate, considering both interim and final analyses, would be maintained at 0.05 \cite{demetsInterimAnalysisAlpha1994}. A formal threshold for the PPoS was not prespecified in the original protocol and the decision on when to perform the final analysis was agreed upon at a consensus meeting that included statisticians and principal investigators of the STHLM3 trial.

\subsection{Results}

In the modelling phase, we modelled the distribution $(T, X \,|\, E=e, A=a)$, where the numeric variable $E$ represents age at baseline (years), and the binary variable $A$ represents the invited ($A=1$) versus non-invited ($A=0$) randomisation group. As enrollment in the trial is complete, modelling of $(E, A)$ is not necessary. To model the distribution of the outcome data, we used piecewise constant cause-specific hazard (PCH) Bayesian models for the two competing events, prostate cancer death ($i=1$) and death from other causes ($i=2$), stratified by invitation group and including age at baseline as a covariate in the model. The cause-specific hazards were piecewise constant on $L_i$ time intervals defined by the boundary knots $q_0=0$ and $q_{L_i}=+\infty$ and by the internal knots $q_i = \{q_{1}, \ldots, q_{L_i-1}\}$\cite{bouazizL0RegularisationEstimation2016}:

\begin{equation*}
\begin{aligned}
\log(\lambda_{ia}(t | e, \beta_{ia}, \gamma_{ia})) &= \beta_{ia1} I_1(t) + \beta_{ia2}  I_2(t) + \ldots + \beta_{iaL_{i}}  I_{L_{i}}(t)  + \gamma_{ia} e \\
&= \sum_{l=1}^{L_{i}} \beta_{ial} I_l(t) + \gamma_{ia} e, 
\end{aligned}
\end{equation*}
where $I_l(t) = \mathbbm{1}(q_{l-1} < t \leq q_l)$\cite{friedmanPiecewiseExponentialModels1982b}. For the cause-specific hazard for PCM, we chose $q_1 = \{2, 3, 4, 5\}$ years, while for the cause-specific hazard for death from other causes we chose $q_2=\{0.5, 1, \ldots, 5.5\}$ years, giving $L_{1}=5$ and $L_{2}=12$ time intervals, respectively. The interim data contained more information on other cause mortality than for PCM (Table \ref{tab:sthlm3-events}), allowing a finer grid for the internal knots. We completed the  models by specifying a lag-1--autoregressive prior on the parameters for the baseline hazard and a weakly informative prior on the parameter for baseline age (Modelling Strategy A in Table \ref{tab-sthlm3-results}). 

In the prediction phase, we simulated the time to event and event type for those men who were still alive at the end of their follow-up period in the interim data ($d_{cens}$), conditioning the predicted event time to be greater than the censoring time. 
%To mimic the fact that the real updated outcome data for a final analysis after 10 years provided by the Swedish National Board of Health and Welfare will cover until the end of 2022, the predicted event times were censored at subject-specific times given by the difference between 2022-12-31 and each subject’s invitation date (time 0). When predicting PPoS for the final analysis with 11 years of follow-up, predictions were censored at the end of 2023, and the same approach was applied for predictions at later time points. After censoring, risk ratios will be evaluated at the last available follow-up time, $FA=\{10.7, 11.7, \ldots,15.7$\}. As enrollment in the STHLM3 trial is complete, the predicted data used in the final analysis $\widetilde{D}_f$ consisted only of $d_0$ and $\widetilde{D}_{cens}$.
To mimic the fact that the outcome data provided yearly by national registries cover the entire calendar year, the final analysis after 10 years will use data until the end of 2022. To this end, the predicted event times were censored at subject-specific times given by the difference between 2022-12-31 and each subject’s invitation date (i.e., time 0), giving a maximum follow-up of 10.7 years. The same approach was applied for predictions at later time points. As enrollment in the STHLM3 trial is complete, the predicted data used in the final analysis $\widetilde{D}_f$ consisted only of $d_0$ and $\widetilde{D}_{cens}$.

\begin{figure}[h]
\centerline{\includegraphics[width=0.8\textwidth]{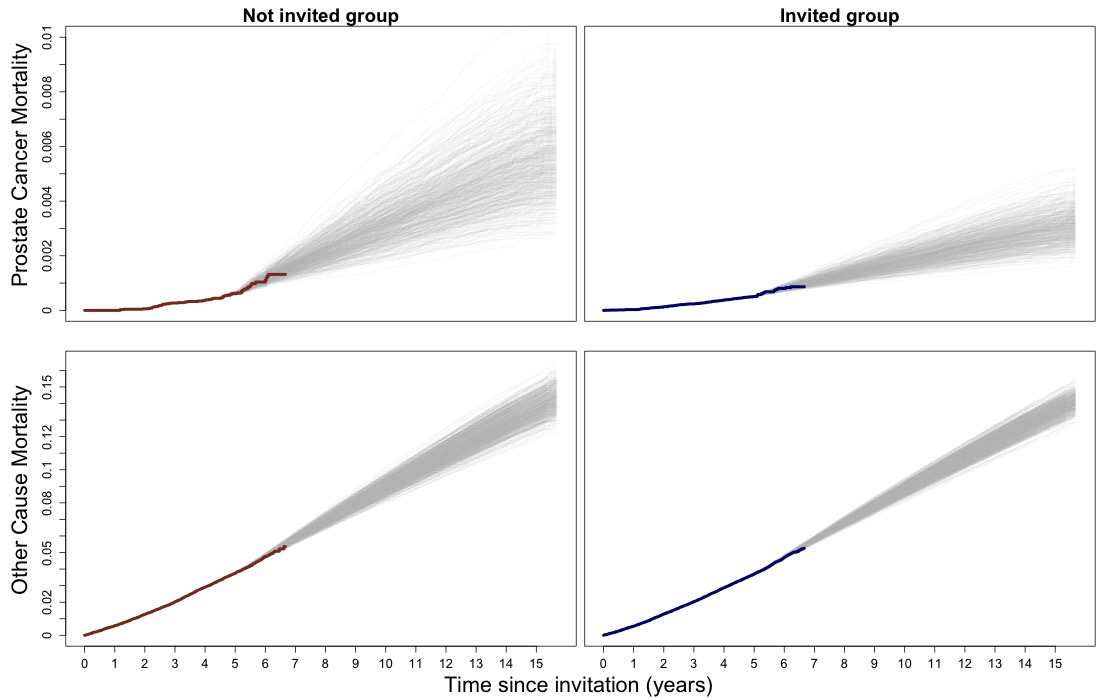}}
\caption{Red and blue thick step lines are the non-parametric crude cumulative risks of death from prostate cancer and of death due to other causes estimated on interim data ($d_0$) for men invited and not invited to participate in the STHLM3 trial, respectively. Grey step lines are a random sample of 500 out of the 2500 non-parametric crude cumulative risk estimates based on interim and predicted data together ($\widetilde D_f$). Predicted data was simulated using Bayesian piecewise constant cause-specific hazard models with 5 and 12 time intervals for death from prostate cancer and death from other causes, respectively (Modelling strategy A).\label{fig2}}
\end{figure}

For each final analysis time point, we repeated $K = 2500$ times the prediction and analysis phase. At each iteration, we estimated the cumulative PCM risks in the two invitation groups (Figure \ref{fig2}) and derived the RR together with the associated p-value at the time of final analysis, corresponding to the largest follow-up time ($FA=\{10.7, 11.7, \ldots,15.7\}$ years). The PPoS was approximated as:
\begin{equation*}
\textrm{PPoS}(FA) \approx \frac{1}{2500}\sum_{k=1}^{2500} \mathbbm{1}(2\cdot\Phi\left(-\left| \frac{\log(\RR(FA)^{(k)})}{ASE(\log(\RR(FA)^{(k)}))} \right|\right) \leq 0.035 \, | \, \tilde\theta^{(k)} ) , % = 0.797.
\end{equation*}
which ranged from 0.783 for the final analysis at 10.7 years of follow-up to 0.892 for the final analysis at 15.7 years (Table \ref{tab-sthlm3-results}, Modelling strategy A).

We performed a sensitivity analysis to assess the extent to which the choices made in the modelling phase could affect the PPoS. In particular, we first explored the consequences of changing the number and location of the internal knots of the PCH models, and then used Bayesian Weibull models for the cause-specific hazards. The specification of the alternative modelling strategies (B--D) is shown in Table \ref{tab-sthlm3-results} together with the corresponding PPoS, which ranged from 0.758 to 0.913 for the earliest time point for the final analysis (10.7 years), and was greater than 0.80 for later time points, regardless of the modelling strategy (Table \ref{tab-sthlm3-results}). Since the PPoS was considered sufficiently high and robust across all modelling strategies for all time points, the statisticians and principal investigators of the STHLM3 trial decided that the PCM risk in the two groups will be compared at 11.7 years after invitation, i.e. with follow-up data until the end of 2023. 
%\deleted{which is expected to provide about 10 years of median follow-up}. % which will be provided by the Swedish National Board of Health and Welfare by the end of 2024.

\begin{table}[b]
\centering
\setlength{\extrarowheight}{1.5pt}
\resizebox{\textwidth}{!}{%
\begin{tabular}{cllcccccc}

\begin{tabular}[c]{@{}c@{}}\textbf{Modelling} \\[-0.4em] \textbf{strategy}\end{tabular} & \textbf{Specification} & \textbf{Priors and hyperpriors} & \begin{tabular}[c]{@{}c@{}}\textbf{PPoS at} \\[-0.4em] \textbf{10.7 years}\end{tabular} & \begin{tabular}[c]{@{}c@{}}\textbf{PPoS at} \\[-0.4em] \textbf{11.7 years}\end{tabular} & \begin{tabular}[c]{@{}c@{}}\textbf{PPoS at} \\[-0.4em] \textbf{12.7 years}\end{tabular} & \begin{tabular}[c]{@{}c@{}}\textbf{PPoS at} \\[-0.4em] \textbf{13.7 years}\end{tabular} & \begin{tabular}[c]{@{}c@{}}\textbf{PPoS at} \\[-0.4em] \textbf{14.7 years}\end{tabular} & \begin{tabular}[c]{@{}c@{}}\textbf{PPoS at} \\[-0.4em] \textbf{15.7 years}\end{tabular} \\ \hline

A  & \begin{tabular}[c]{@{}l@{}}
PCH model \\ $L_1 = 5$ ,  $L_2 = 12$ \\ 
$q_1 = \{2, 3, 4, 5\}$ \\
$q_2 = \{0.5, 1, \ldots, 5.5\}$ 
\end{tabular}       
& \begin{tabular}[c]{@{}l@{}}
$\beta_{ia1}  \sim \textrm{Normal}(-10,20)$
\\ $\beta_{ial} - \beta_{ia(l-1)} \sim  \textrm{Normal}(0, \tau_{ia})$ for $l = 2, \ldots,  L_i$  
\\ $\gamma_{ia} \sim \textrm{Normal}(0, \sqrt{0.5})$ 
\\ $\tau_{ia} \sim \textrm{Exponential}(1)$ \\ \end{tabular} 
& 0.783 & 0.826 & 0.856 & 0.864 & 0.885 & 0.892 \\ \hline

B  & \begin{tabular}[c]{@{}l@{}}
PCH model \\ $L_1 = L_2 = 3$ \\ 
$q_1 = q_2 = \{2, 4\}$ 
\end{tabular}       
& \begin{tabular}[c]{@{}l@{}}
$\beta_{ia1}  \sim \textrm{Normal}(-10,20)$
\\ $\beta_{ial} - \beta_{ia(l-1)} \sim  \textrm{Normal}(0, \tau_{ia})$ for $l = 2, 3$  
\\ $\gamma_{ia} \sim \textrm{Normal}(0, \sqrt{0.5})$ 
\\ $\tau_{ia} \sim \textrm{Exponential}(1)$ \\ \end{tabular} 
& 0.758 & 0.796 & 0.832 & 0.868 & 0.877 & 0.892 \\ \hline

C  & \begin{tabular}[c]{@{}l@{}}
PCH model \\ $L_1 = L_2 = 5$ \\ 
$q_1 = q_2 = \{2, 3, 4, 5\}$ \end{tabular}       
& \begin{tabular}[c]{@{}l@{}}
$\beta_{ia1}  \sim \textrm{Normal}(-10,20)$
\\ $\beta_{ial} - \beta_{ia(l-1)} \sim  \textrm{Normal}(0, \tau_{ia})$ for $l = 2, \ldots, 5$  
\\ $\gamma_{ia} \sim \textrm{Normal}(0, \sqrt{0.5})$ 
\\ $\tau_{ia} \sim \textrm{Exponential}(1)$ \\ \end{tabular} 
& 0.783 & 0.825 & 0.850 & 0.863 & 0.884 & 0.894 \\ \hline

D & \begin{tabular}[c]{@{}l@{}}
Weibull model \\
$\log(u_{ia}) = \alpha_{ia} + \gamma_{ia}e$
\end{tabular}                                                                                  
& \begin{tabular}[c]{@{}l@{}}
$\alpha_{ia} \sim \textrm{Normal}(-10, 20)$ 
\\ $\gamma_{ia} \sim \textrm{Normal}(0, \sqrt{0.5})$
\\ $\nu_{ia} \sim \textrm{Exponential}(1)$\\ \end{tabular}                                                                                                             
& 0.913 & 0.946 & 0.958 & 0.967 & 0.976 & 0.980 \\ \hline                   

\end{tabular}%
}
\caption{Modelling strategies used in the modelling stage of the STHLM3 trial PPoS analysis. PPoS at different time points for the final analysis are reported. Subscript $i$ is the event type: $i=1$ death due to prostate cancer, $i=2$ death due to other causes. Subscript $a$ identifies the randomisation group: $a=0$ not invited, $a=1$ invited. PCH: piecewise constant cause-specific hazard. }
\label{tab-sthlm3-results}
\end{table}

\clearpage

\section{Discussion}\label{sec5}

In this paper, we have described a simulation-based approach to compute the PPoS for interim monitoring of clinical trials when the survival outcome is subject to competing events. 

In the modelling phase, we have proposed the use of Bayesian cause-specific hazard models to model the joint distribution of time to event and event type given covariates. However, alternative approaches are possible and include multivariate latent failure time models \cite{mengSimulatingTimetoeventData2023}, vertical models\cite{nicolaieVerticalModelingPattern2010}, and arguably cumbersome models based on the subdistribution hazards \cite{bonnevilleWhyYouShould2024}. We have chosen cause-specific hazard-based models because they provide a convenient and flexible modelling framework \cite{crowtherGeneralFrameworkParametric2014}. In addition, as the cause-specific hazards fully specify the joint distribution $(T,X)$ based on observable quantities alone, unverifiable assumptions about the dependence structure of the latent event times can be avoided. Different models can be specified to explore how sensitive the PPoS is to the modelling assumptions, as we did in Section \ref{sec4}. The use of baseline covariates other than randomisation arm can in principle improve the prediction of the outcome data \cite{avalos-pachecoValidationPredictiveAnalyses2023}. This comes at the cost of an increased complexity due to the need to model their joint distribution, if new patients' baseline data are to be simulated. A simple approach that however ignores parameter uncertainty is to sample with replacement from the baseline covariates of the interim data, effectively sampling from their empirical joint distribution. 

The specification of the prior distributions represents a common concern with any approach based on Bayesian methodology, and the one proposed here is no exception. Weakly informative priors for the parameters of the model for $(T,X)$ can be specified so that the predictions can be considered to be based on interim data only\cite{spiegelhalterMonitoringClinicalTrials1986}. Informative priors may be otherwise specified to incorporate external information about the data-generating process, including the treatment effect\cite{rufibachSequentiallyUpdatingLikelihood2016}. For PPoS sensitivity analyses, informative priors may be used to anticipate or explore the consequences of changes in the population or in the treatment effect over time\cite{savilleConditionalPowerHow2023}, or to counterbalance the prior opinions of someone who would doubt the observed results \cite{spiegelhalterBayesianApproachesRandomized1994}, as we did in Section \ref{sec3}. Ultimately, the choice of the priors should be aligned with the objectives of the analysis. If necessary, the choice of the prior distributions can be discussed and agreed upon with the regulators at the time of trial design, when the operating characteristics of the trial are assessed. Finally, it should be noted that the priors used in the modelling phase do not need to be the same as those used in the analysis phase. 

As the proposed approach is simulation-based, it can be computationally intensive, especially as the complexity of the prediction and analysis phase increases. Depending on the specific Bayesian models used in the modelling phase, predicting survival times can be a time-consuming task. This is the case when the cumulative distribution function for the distribution of $T$ cannot be inverted analytically (as with cause-specific hazard Weibull models) or when it has to be evaluated by numerical integration \cite{crowtherSimulatingBiologicallyPlausible2013}. The analysis phase can also be computationally demanding, especially when Bayesian methods are used. For reference, computing the PPoS for the I-SPY COVID example took approximately 9 minutes on a high-performance computing cluster using 40 cores, while for the larger STHLM3 trial it took over 10 hours. The computational burden increases at the design stage, when the trial's operating characteristics need to be assessed for a possibly large number of scenarios. The number of simulations $K$ can be guided not only by time and computational constraints, but also by the accuracy with which the PPoS needs be approximated \cite{koehlerAssessmentMonteCarlo2009}. For example, our choice of $K=2500$ was motivated by setting the upper bound on the standard error of the PPoS to 0.01. On the other hand, our approach offers full flexibility in the modelling and analysis phase and does not require resorting to asymptotic approximations \cite{marion_predictive_2024} or to the use of conjugate priors \cite{berryBayesianAdaptiveMethods2010}. In particular, any statistical Bayesian or frequentist method for competing event data can be used in the analysis phase, including methods for the number of life years lost due to the different competing events \cite{andersenDecompositionNumberLife2013} or for joint inference on the crude cumulative risks of multiple competing events \cite{wenSimultaneousHypothesisTesting2023}.

Our work was motivated by two randomised clinical trials, both of which had a primary survival outcome with death as a competing event. In the I-SPY COVID trial, the PPoS analysis complemented the prespecified efficacy analysis, which together led the DMC to recommend stopping the randomisation in the backbone-plus-cyclosporine arm due to futility. In the STHLM3 trial, the PPoS analysis was used to decide to perform the final analysis to compare the long-term PCM mortality risk between men randomly invited or not invited to prostate cancer screening using follow-up data that cover until end of 2023. %at 11.7 years after invitation.

The use of Bayesian PPoS for interim monitoring of clinical trials is naturally attractive because they are easy to interpret for trial stakeholders, including physicians, patients, and DMC board members, and because they directly address a relevant question: how likely is it that the trial will achieve its objective, given the current data, if it continues to its end? Importantly, this predictive question is addressed taking into account two sources of variability: the variability in the data not yet observed at interim monitoring and the variability in the parameter estimates for the data-generating process. At the same time, the calculation of the PPoS involves a number of assumptions that may lead to an inaccurate prediction of the success probability if they do not hold true\cite{avalos-pachecoValidationPredictiveAnalyses2023, savilleConditionalPowerHow2023}. As we have done in our motivational examples, we recommend performing sensitivity analyses to evaluate how different assumptions may affect the PPoS\cite{rufibachSequentiallyUpdatingLikelihood2016}.

In conclusion, our work extends the existing literature on the PPoS by proposing a simulation-based approach that allows the PPoS to be used for interim monitoring of clinical trials in the common setting where the trial outcome is subject to competing events.

%\vspace*{12pt}
\bmsection*{Data/Code Availability }
%\vspace*{12pt}
The R code that was used to analyse the data and some synthetic data can be obtained from \url{https://github.com/cmicoli/PPoS-CompetingRisks}.

%\backmatter
%\bmsection*{Author contributions}

%???? This is an author contribution text. This is an author contribution text. This is an author contribution text. This is an author contribution text. This is an author contribution text.

\bmsection*{I-SPY COVID Funding and Acknowledgements:}
This work was funded by COVID R\&D Consortium, Allergan, Amgen Inc., Takeda Pharmaceutical Company, Ingenus Pharmaceuticals LLC, Implicit Bioscience, Johnson \& Johnson, Pfizer Inc., Roche/Genentech, Apotex Inc., FAST Grant from Emergent Venture George Mason University, The DoD Defense Threat Reduction Agency (DTRA), The Department of Health and Human Services Biomedical Advanced Research and Development Authority (BARDA), and The Grove Foundation.
\bmsection*{STHLM3 Funding and Acknowledgements:}
This work was funded by the Swedish Cancer Society (Cancerfonden), the Swedish Research Council (Vetenskapsrådet), Region Stockholm, and Svenska Druidorden, Åke Wibergs Stiftelse, the Swedish e-Science Research Center, the Karolinska Institutet, and Prostatacancerförbundet. The STHLM3 trial was funded by Region Stockholm.

%\bmsection*{Financial disclosure}
%????? None reported.

\bmsection*{Conflict of interest}
ME reports four pending prostate cancer diagnostic-related patents: method for indicating a presence or non-presence of aggressive prostate cancer (WO2013EP7425920131120); prognostic method for individuals with prostate cancer (WO2013EP7427020131120); method for indicating a presence of prostate cancer in individuals with particular characteritics (WO2018EP5247320180201); and method for indicating the presence or non-presence of prostate cancer (WO2013SE5055420130516). The Karolinska Institutet collaborates with A3P Biomedical in developing the technology for the Stockholm3 test. ME owns shares in A3P Biomedical. All other authors declare no competing interests.

\bibliography{wileyNJD-AMA}

%\bmsection*{Supporting information}
%Additional supporting information may be found in the
%online version of the article at the publisher’s website.

\newpage

\appendix

\textbf{COVID Executive Committee} \\
Laura J. Esserman; Carolyn S. Calfee; Michael A. Matthay; Kathleen D. Liu; D. Clark Files; Martin Eklund; Karl W. Thomas.
\\
\textbf{ARDS Executive Committee}\\
Laura J. Esserman; Kathleen D. Liu; D. Clark Files; Kevin Gibbs; Derek W. Russell; Martin Eklund; Karl W. Thomas. 

\begin{table}[!ht]
\centering
\begin{tabular}{|l|l|p{10.5cm}|}
\hline
\textbf{Site} & \textbf{Principal Investigators (PI)} & \textbf{Affiliation} \\
\hline
Columbia NYC & Jeremy R. Beitler, MD, MPH & Centre for Acute Respiratory Failure, Columbia University \\
\hline
Emory & Sara C. Auld, MD & Department of Medicine, Emory University \\
\hline
Georgetown & Nathan Cobb, MD & Georgetown University Medical Centre \\
\hline
Hoag (Irvine) & Philip Robinson, MD & Hoag Memorial Hospital Presbyterian Centre for Research and Education \\
\hline
Hoag (Newport Beach) & Philip Robinson, MD & Hoag Memorial Hospital Presbyterian Centre for Research and Education \\
\hline
KP LAMC & Kenneth K. Wei, MD & Division of Pulmonary and Critical Care Medicine, Kaiser Permanente Los Angeles Medical Center \\
\hline
Logan Health & Timothy Obermiller, MD & Critical Care Medicine, Logan Health Research Institute \\
\hline
Long Beach Memorial & Fady A. Youssef, MD & Department of Internal Medicine, Pulmonary Division, Long Beach Memorial Medical Centre \\
\hline
Main Line Health & Eliot Friedman, MD & Department of Medicine, Main Line Health \\
\hline
Mercy Hospital & Sandya Samavedam, MD & Critical Care Medicine, Mercy Hospital \\
\hline
Northwestern & Richard G. Wunderink, MD & Department of Medicine, Pulmonary and Critical Care Division, Northwestern University Feinberg School of Medicine \\
\hline
Sanford Health & Paul Berger, DO & Department of Critical Care, Sanford Health \\
\hline
Spectrum Health & Malik M. Khan, MD & Pulmonary and Critical Care, Spectrum Health \\
\hline
Stamford Health & Michael Bernstein, MD & Pulmonary Medicine, Stamford Health \\
\hline
U Penn & Nuala J. Meyer, MD & Department of Medicine, University of Pennsylvania Perelman School of Medicine \\
\hline
UAB & \begin{tabular}[c]{@{}l@{}}Derek W. Russell, MD \\ Sheetal Gandotra, MD\end{tabular} & \begin{tabular}[c]{@{}l@{}}Pulmonary, Allergy, and Critical Care Medicine, University of Alabama, Birmingham \\ Pulmonary, Allergy, and Critical Care Medicine, University of Alabama, Birmingham\end{tabular} \\
\hline
UC Davis & Timothy Albertson, MD & Division of Pulmonary, Critical Care and Sleep Medicine, University of California Davis \\
\hline
UC Irvine & Richard A. Lee, MD & Division of Pulmonary Diseases and Critical Care Medicine, University of California \\
\hline
UCSF & \begin{tabular}[c]{@{}l@{}}Kathleen D. Liu, MD, PhD \\ Laura J. Esserman, MD \\ Carolyn S. Calfee, MD \\ Michael A. Matthay, MD\end{tabular} & \begin{tabular}[c]{@{}l@{}}Divisions of Nephrology and Critical Care Medicine, University of California San Francisco \\ Department of Surgery \\ Division of Pulmonary, Department of Medicine, Critical Care, Allergy and Sleep Medicine \\ Division of Pulmonary, Department of Medicine, Critical Care, Allergy and Sleep Medicine\end{tabular} \\
\hline
University of Colorado & Ellen L. Burnham, MD & Department of Medicine, University of Colorado \\
\hline
University of Miami & \begin{tabular}[c]{@{}l@{}}Christopher P. Jordan, MD \\ Daniel H. Kett, MD\end{tabular} & \begin{tabular}[c]{@{}l@{}}Department of Pulmonary Care and Sleep Medicine \\ Department: Medicine, Division: Pulmonary, Critical Care and Sleep Medicine\end{tabular} \\
\hline
University of Michigan & Robert Hyzy, MD & Division of Pulmonary and Critical Care Medicine, Department of Medicine, University of Michigan \\
\hline
USC & Santhi Kumar, MD & Division of Pulmonary, Critical Care and Sleep Medicine, Keck School of Medicine, USC \\
\hline
Wake Forest & \begin{tabular}[c]{@{}l@{}}D. Clark Files, MD \\ Kevin W. Gibbs, MD\end{tabular} & \begin{tabular}[c]{@{}l@{}}Department of Internal Medicine, Wake Forest University \\ Pulmonary Critical Care, Allergy and Immunologic Disease\end{tabular} \\
\hline
West Virginia University & Jeremiah Hayanga, MD & Department of Cardiovascular and Thoracic Surgery, Heart and Vascular Institute, West Virginia University \\
\hline
Yale & Jonathan L. Koff, MD & Section of Pulmonary, Critical Care, and Sleep Medicine, Yale University \\
\hline
\end{tabular}
\caption{List of Sites, Principal Investigators, and Affiliations of the I-SPY COVID Consortium.}
\end{table}

\end{document}